\documentclass[10pt,twocolumn]{article} 
\usepackage{arxivStyle}
\usepackage{times}
\usepackage{graphicx}
\usepackage{amssymb}
\usepackage{url,hyperref}

\begin{document}

\title{DeeperHistReg: Robust Whole Slide Images Registration Framework}

\author{Marek Wodzinski$^{1,2,*}$, Niccol\`{o} Marini$^{1}$, Manfredo Atzori$^{1,3}$, Henning M\"{u}ller$^{1,4}$
\\
\\
$^{1}$Information Systems Institute, University of Applied Sciences Western Switzerland, Sierre, Switzerland \\ 
$^{2}$Department of Measurement and Electronics, AGH University of Kraków, Krakow, Poland \\
$^{3}$Department of Neuroscience, University of Padova, Padova, Italy \\
$^{4}$Medical Faculty, University of Geneva, Geneva, Switzerland \\
$^{*}$Correspondence: {wodzinski@agh.edu.pl}
}

\maketitle
\thispagestyle{empty}

\begin{abstract}
DeeperHistReg is a software framework dedicated to registering whole slide images (WSIs) acquired using multiple stains. It allows one to perform the preprocessing, initial alignment, and nonrigid registration of WSIs acquired using multiple stains (e.g. hematoxylin \& eosin, immunochemistry). The framework implements several state-of-the-art registration algorithms and provides an interface to operate on arbitrary resolution of the WSIs (up to 200k x 200k). The framework is extensible and new algorithms can be easily integrated by other researchers. The framework is available both as a PyPI package and as a Docker container.

\end{abstract}

\begin{figure*}[htb]
	\centering
    \includegraphics[scale=0.55]{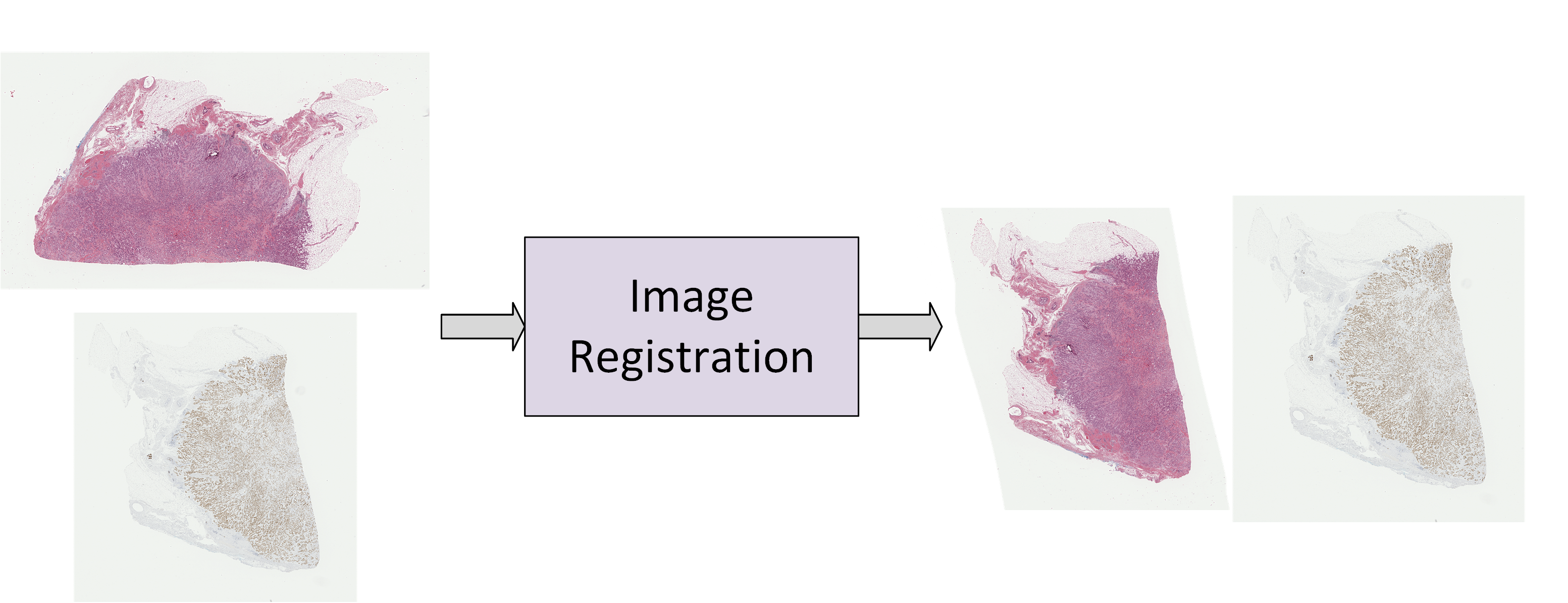}
    \caption{Overview of the WSI registration. The goal is to align the input images to the same geometrical space.}   
    \label{fig:overview}
\end{figure*}

\section{Motivation and significance}

The automatic and robust registration of whole slide images (WSIs) is an important task in the digital pathology pipeline~\ref{fig:overview}. There are three major applications of WSI registration: (i) to fuse information from slides acquired using hematoxylin \& eosin (H\&E) and immunochemistry (IHC) stainings, (ii) to perform 3-D tissue reconstruction using the consecutive slides, (iii) to transfer annotations from one annotated slide to the remaining slides of a given tissue~\cite{borovec2020anhir}. All these applications are important from the perspective of diagnostic quality, as well as the appropriate use of resources. The importance of the task led researchers around the world to organize open scientific challenges dedicated to WSI registration, e.g. ANHIR~\cite{borovec2020anhir} and ACROBAT~\cite{weitz2023acrobat}.

There are numerous scientific contributions addressing the WSI registration~\cite{lotz2015patch,marzahl2021robust,gatenbee2023virtual,solorzano2018whole,lotz2021comparison2,awan2023deep,wodzinski2021multistep,wodzinski2021deephistreg}. Nevertheless, their main limitation is related to the source code availability, quality, reproducibility, and practicality, with the exception of the VALIS software that is fully operational~\cite{gatenbee2023virtual}. Most of the researchers do not share the source code. Moreover, even if the source code is released, the convenient interface to perform the registration and save the results using the resolution and format desired by pathologists is unavailable. Based on the experience from our scientific contributions~\cite{wodzinski2021multistep,wodzinski2021deephistreg,wodzinski2024}, the interest of other researchers in having ready-to-use WSI registration is significant. To date, more than 80 researchers worldwide requested the pretrained models and support with running the aforementioned methods. Therefore, we decided to release an updated, ready-to-use software library that allows one to perform fully automatic and robust WSI registration.

\textbf{Contribution: } The released software package enables one to automatically register any OpenSlide-compatible WSIs and save them in pyramid TIFF. We support both initial affine alignment, as well as detailed nonrigid registration. The library does not require any dedicated fine-tuning to a particular dataset and contains state-of-the-art algorithms that took high places in the ANHIR and ACROBAT competitions. We plan to maintain support and implement additional functionalities if requested by the library users. 

\section{Software description}

\subsection{Overview \& Software capabilities}

The software aims to provide a convenient interface to load a pair of WSIs, perform the registration, and save the transformed moving image aligned to the fixed image. Optionally, it is possible to save the calculated displacement field together with the parameters required to transform the annotations: (i) landmarks, (ii) shapes, (iii) segmentation masks. Examples of how to use the library to transfer the annotations are available in the associated repository.

The user configures the pipeline using a JSON file with the desired registration steps and algorithms. The configuration file should contain parameters required by the desired preprocessing, initial alignment, and nonrigid registration methods. Several pre-defined configurations are available in the method repository.

\subsection{Software architecture}

\begin{figure*}[!htb]
	\centering
    \includegraphics[scale=0.80]{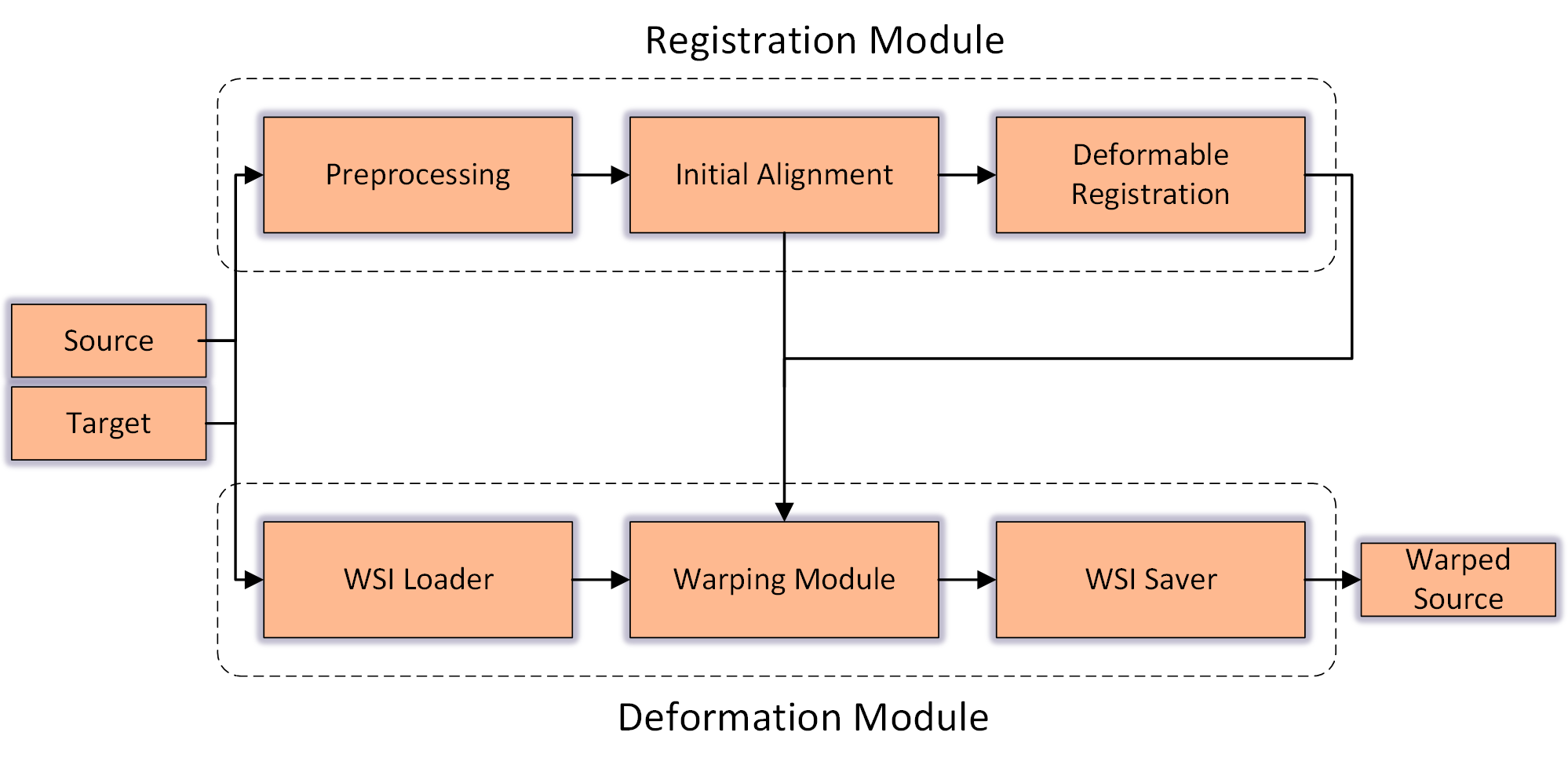}
    \caption{The DeeperHistReg processing pipeline.}   
    \label{fig:pipeline}
\end{figure*}

The software consists of three main modules: (i) an input/output (IO) module, (ii) a registration module, and (iii) a deformation module. The software structure and the processing pipeline are shown in Figure~\ref{fig:pipeline}.

\begin{figure*}[!htb]
	\centering
    \includegraphics[scale=0.75]{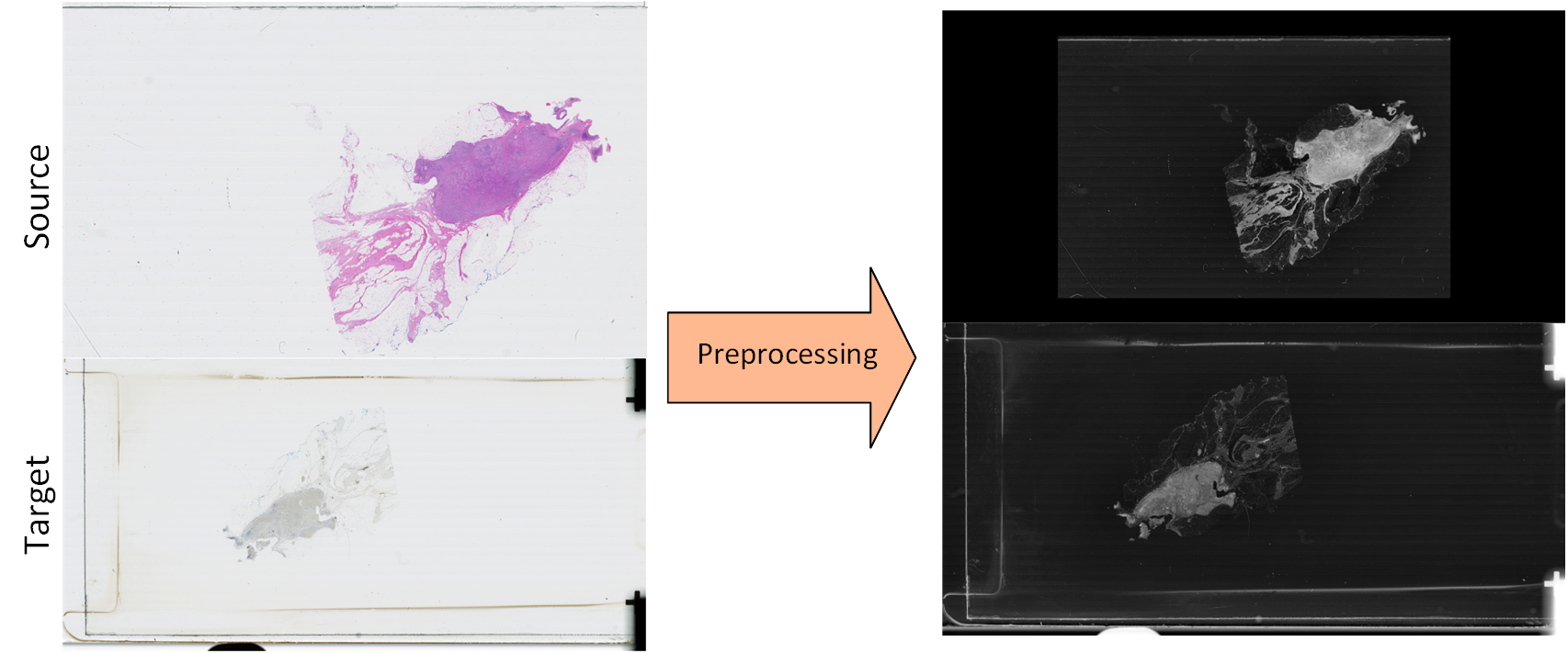}
    \caption{Visualization of WSIs preproccesing using slides from the ACROBAT dataset. The images are initially resampled, padded and converted to grayscale.}   
    \label{fig:preprocessing}
\end{figure*}

The IO module provides an interface to conveniently load and save a pair of WSIs at any chosen pyramid level. The module is based on PyVips~\cite{pyvips} and is able to load any OpenSlide-compatible format. The transformed WSIs are saved to pyramidal TIFF. Thanks to PyVips, it is possible to load and save WSIs at arbitrary resolution (verified up to 200k x 200k). The IO module also implements basic operations like padding to the same shape and resampling by a chosen factor.

The registration module consists of several submodules: (i) the preprocessing module, (ii) the initial alignment module, (iii) the deformable registration module. The preprocessing module is responsible for initial intensity correction and conversion to the representation desired by the following registration methods. The task of the initial alignment module is to automatically calculate the initial linear transformation (rigid, affine, or projective). The task is considerably difficult for consecutive WSI acquired using multiple stains because no assumptions about the initial orientation can be made. The deformable module calculates the final nonrigid transformation that can be based on e.g. B-Splines or dense displacement fields.

The deformation module is based on the PyVips library. It converts the calculated displacement field to a PyVips Image and applies patch-based warping to the WSI. The warping is performed iteratively with a pre-defined patch size that makes it possible to transform WSI at any desired resolution. As a result, it is possible to efficiently transform high-resolution images using the displacement field calculated by the registration module. Moreover, the deformation module allows one to transform the associated segmentation masks or landmarks to make the software library useful for annotation transfer.

\subsection{Availability \& Usage}

The source code is available on GitHub with the CC-BY-SA license~\cite{source_code}, with the exception of several, optional, deep learning models with different licenses (described in the repository). The instructions on how to install and use the library are available in the repository. The library is available for installation in the PyPI package index. Moreover, a Docker container that contains all the required dependencies is downloadable directly from the GitHub repository.

\section{Illustrative examples}

\subsection{Preprocessing}

The library can be used for WSI preprocessing, e.g. by initial resampling, padding, and converting to grayscale (required by the majority of the objective functions used for image registration). An example of WSIs from the ACROBAT dataset~\cite{borovec2020anhir} before and after the preprocessing is presented in Figure~\ref{fig:preprocessing}.

\begin{figure*}[!htb]
	\centering
    \includegraphics[scale=0.70]{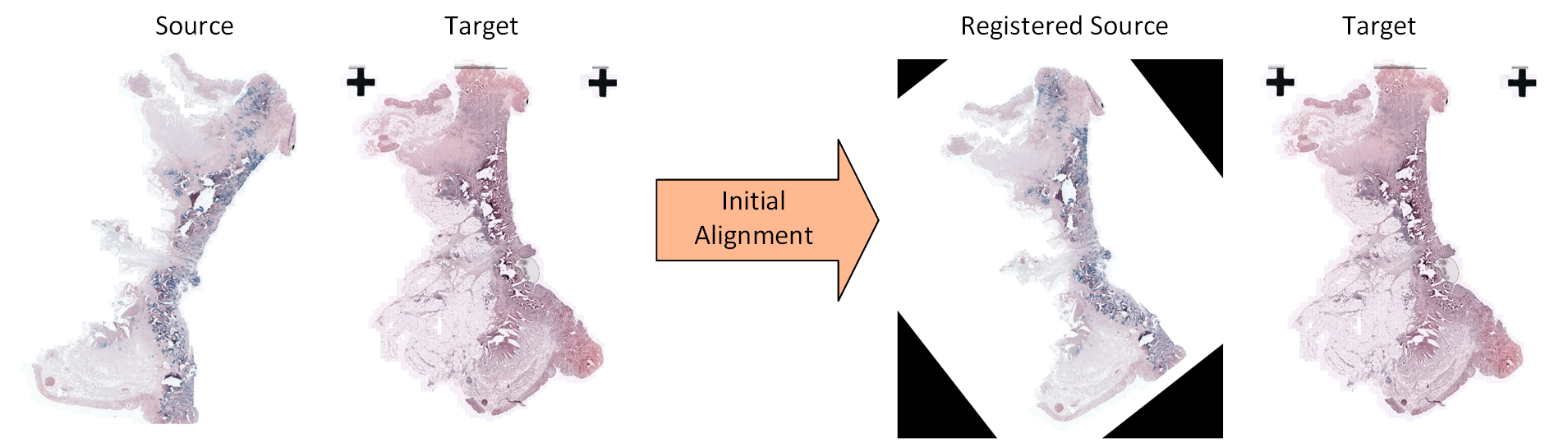}
    \caption{Exemplary WSI initial alignment result using images from the ANHIR dataset. The images are initially aligned to the same geometrical space, to be further improved by the nonrigid registration.}   
    \label{fig:initial_alignment}
\end{figure*}

\begin{figure*}
	\centering
    \includegraphics[scale=0.95]{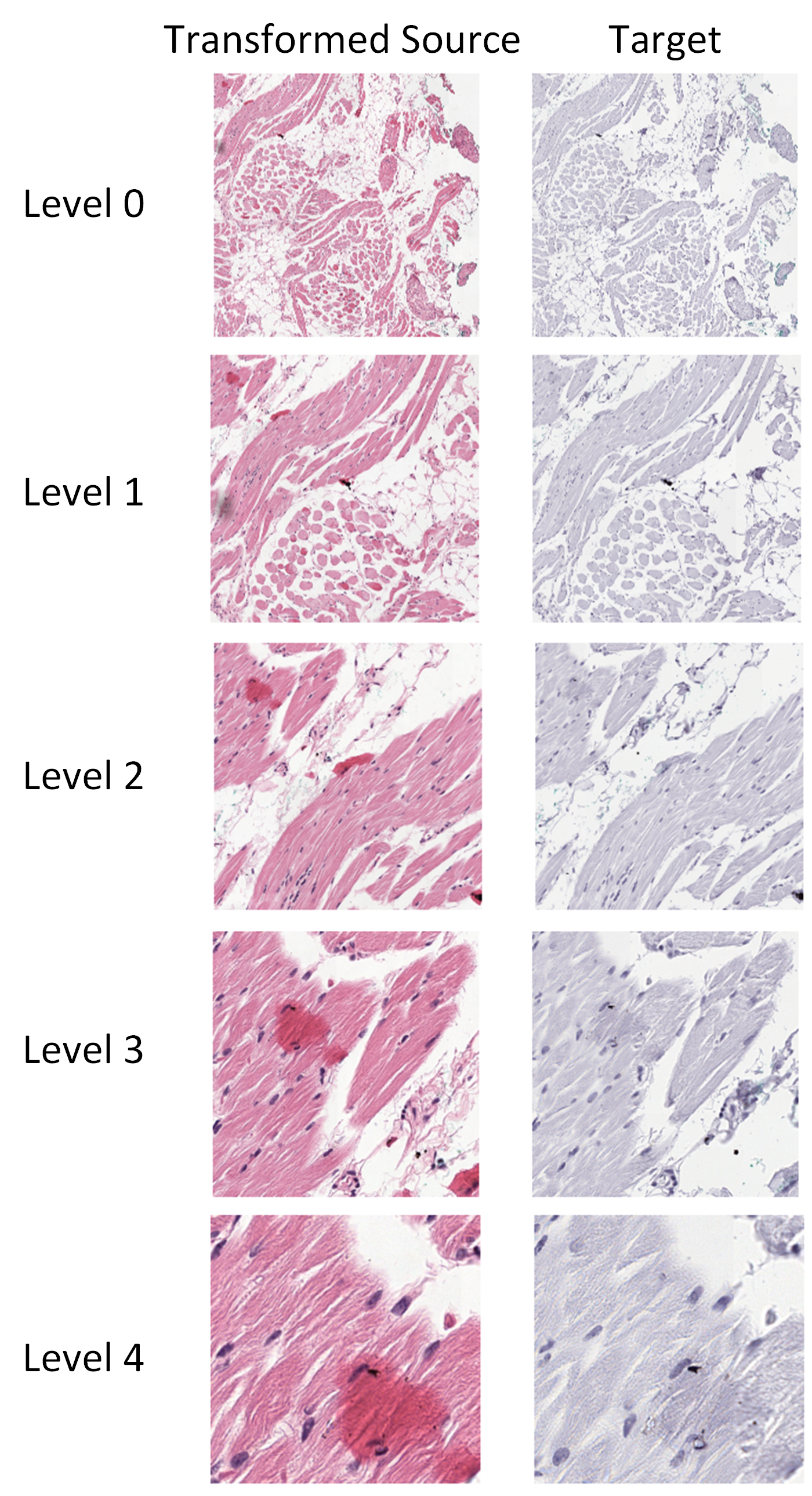}
    \caption{Exemplary results of the deformable registration at various magnification levels using a re-stained H\&E and PHH3 slides from the HyReCo database. Please note the spatial agreement even at the cell level.}   
    \label{fig:nonrigid_registration}
\end{figure*}

\subsection{Affine Registration}

The library is useful for performing the initial affine registration of WSIs. The affine registration may be useful for e.g. 3-D reconstruction from numerous consecutive slices. Exemplary results of the initial alignment using WSIs from the ANHIR dataset~\cite{weitz2022acrobat} are presented in Figure~\ref{fig:initial_alignment}.

\subsection{Nonrigid Registration}

The nonrigid registration is especially important for registering highly deformed consecutive or restained slices. A potential use-case of such registration is to perform annotation transfer or pre-register two or more stains to improve deep learning-based classification of segmentation methods. Exemplary results of the deformable registration using restained WSIs from the HyReCo dataset~\cite{lotz2021comparison} are presented in Figure~\ref{fig:nonrigid_registration}.

\subsection{Computational details}

The computational time required to run the registration depends on the configuration and the WSIs resolution. For example, the full registration of a 40k x 40k ACROBAT slides takes about: (i) 40 seconds for the registration, (ii) 20 seconds for the final warping. On the other hand, the registration of restained HyReCo slide (100k x 100k) requires about: (i) 2 minutes for the registration, (ii) 6 minutes for the final warping (at the highest resolution level), using NVIDIA A6000 and Intel Xeon Gold 5518 with 16 cores. The registration and warping times are among the most effective ones and are widely acceptable in the field of digital pathology.

\section{Impact and conclusions}

The DeeperHistReg may have a significant impact on the field of digital pathology. Automatic, robust, and efficient registration of WSIs and other microscopy images is beneficial for several applications: (i) content-based image retrieval, (ii) annotation transfer, (iii) 3-D reconstruction of consecutive slides, (iv) improving multimodal classification/segmentation algorithms. All these applications are important for AI-based digital pathology and may lead to breakthroughs in cancer diagnosis.

The algorithms available in the DeeperHistReg library are among best-performing ones worldwide, both from qualitative and quantitative points of view. The methods were on the podium of the IEEE ISBI ANHIR challenge and won the ACROBAT challenge organized during the MICCAI 2023 conference. Considering the huge interest from the scientific community in the WSI registration - the ANHIR challenge article has already almost 100 citations and we got more than 80 requests for releasing the DeepHistReg pretrained weights - releasing the library will have a significant impact on the field.

The future updates of DeeperHistReg will include direct support for the BioFormats library to make it useful for various other microscopy images. Moreover, additional nonrigid registration methods, based on the most recent advances in deep learning, will be introduced to the library to make the results as accurate as possible. Finally, we plan to release another library dedicated to WSI 3-D reconstruction, built on top of the DeeperHistReg source code.

\section*{Acknowledgements}
%\label{}

This project has received funding from the Innovative Medicines Initiative 2 Joint Undertaking under grant agreement No 945358. This Joint Undertaking receives support from the European Union's Horizon 2020 research and innovation program and EFPIA, Belgium (www.imi.europe.eu). The research reflects only the author's view and the Joint Undertaking is not responsible for any use that may be made of the information it contains. Additionally, the research was supported in part by PLGrid Infrastructure. We gratefully acknowledge Poland’s high-performance computing infrastructure PLGrid (HPC Centers: ACK Cyfronet AGH) for providing computer facilities and support within computational grant no. PLG/2023/016239.

\section*{Declaration of competing interest}
%\label{}

The authors declare that they have no known competing financial interests or personal relationships that could have appeared
to influence the work reported in this paper.

\section*{Data availability}
%\label{}

All data used in the article is openly available.

\bibliographystyle{abbrv}
\bibliography{main}
\end{document}